\documentclass[12pt]{article}

\usepackage{amsmath,amsfonts,epsfig,color,latexsym}

\newcommand{\be}{\begin{equation}}
\newcommand{\ee}{\end{equation}}
\newcommand{\bea}{\begin{eqnarray}}
\newcommand{\eea}{\end{eqnarray}}

\begin{document}

\begin{flushright}
hep-th/0501068\\
BROWN-HET-1439
\end{flushright}
\vskip.5in

\begin{center}

{\LARGE\bf The RHIC fireball as a dual black hole }
\vskip 1in
\centerline{\Large Horatiu Nastase}
\vskip .5in

\end{center}
\centerline{\large Brown University}
\centerline{\large Providence, RI, 02912, USA}

\vskip 1in

\begin{abstract}

{\large We argue that the fireball observed at RHIC is (the analog of)
a dual black hole. In previous works, we have argued that the large 
$s$ behaviour of the total QCD cross section
is due to production of dual black holes, and that in the QCD effective 
field theory  it corresponds to a nonlinear soliton of the pion field. 
Now we argue that the RHIC fireball is this soliton. We calculate the 
soliton (black hole) temperature, and get $T=4a <m_{\pi}>/\pi$, with $a$ 
a nonperturbative constant. For $a=1$, we get $175.76 MeV$, compared to 
the experimental value of the fireball ``freeze-out'' of 
about $176 MeV$. The 
observed $\eta/ s$ for the fireball is close to the dual value of $1/4\pi$. 
The ``Color Glass Condensate'' (CGC) state at the core of the fireball 
is the pion field soliton, dual to the interior of the black hole. 
The main interaction between particles in the CGC is a Coulomb potential,
due to short range pion exchange, dual to gravitational interaction 
inside the black hole, deconfining quarks and gluons. 
Thus RHIC is in a certain sense a string theory testing machine, analyzing
the formation and decay of dual black holes, and giving information about 
the black hole interior.    
}

\end{abstract}

\newpage

The study of high energy scattering in QCD via AdS-CFT \cite{malda}
was started by the work of Polchinski and Strassler \cite{ps}. They 
argued that high energy scattering in QCD with momentum p and 
wavefunction $e^{ipx}$ corresponds to scattering in a space $AdS_5\times 
X_5$, 
\be
ds^2= \frac{\bar{r}^2}{R^2} d\vec{x}^2 + \frac{R^2}{\bar{r}^2}d\bar{r}^2 + 
R^2 ds_X^2= e^{-2y/R}d\vec{x}^2 + dy^2 + R^2 ds_X^2
\ee
with $AdS_5$ cut off in the IR (at small $\bar{r}$, or large y), and 
local AdS momentum $\tilde{p}_{\mu}= (R/\bar{r}) p_{\mu}$ and wavefunction
$e^{ipx}\psi(\bar{r}, \Omega_5)$. 
The cut off corresponds to our ignorance of the 
IR modification of the gravity dual and is represented by an IR 
brane, giving a RS model \cite{rs}.

Later, Giddings \cite{gid} argued that one will start to create black holes
in AdS when one reaches the Planck scale $M_P$, and this corresponds in 
gauge theory to a power law behaviour of the total cross section
above $\hat{M}_P= N^{1/4}\Lambda_{QCD}$, and 
that the gauge theory 
maximal  Froissart behaviour will onset when the size of the 
black holes reaches the AdS size, and the black hole is on the IR brane.

In \cite{kntwo}, this picture was made precise. The AdS scattering was 
modelled by scattering of Aichelburg-Sexl shockwaves \cite{as,thooft}, moving 
in the AdS background \cite{nastase}. The shockwaves in flat space are
\be
ds^2 = 2dx^+ dx^- +(dx^+)^2 \Phi (x^i) \delta (x^+) +d\vec{x}^2
\label{shock}
\ee
where the function $\Phi $ satisfies the Poisson equation
\be
\Delta_{D-2} \Phi (x^i) = -16 \pi G p \delta ^{D-2}(x^i)
\label{poisson}
\ee
In curved space, one just puts the above in the background, and the 
laplacean $\Delta$ is also in the background. It was found that indeed, 
the scattering produces black holes, and then 
for flat space scattering one obtains a power law 
for the gauge theory cross section, $\sigma_{gauge}\sim s^{1/(D-3)}$, and 
for scattering on the IR brane one obtains maximal Froissart behaviour
$\sigma_{gauge}\sim 1/M^2 ln^2 s$. 

In \cite{knthree} we have seen that this picture maps exactly to the 
1952 effective field theory picture of Heisenberg \cite{heis}
for the saturation of 
the Froissart bound, of collisions of shockwaves distributions of the 
pion field. To be exact, the dual picture refers to the case when there is 
no pion (almost Golstone boson), but only a lightest glueball (dual to the
lightest KK gravity mode). But there is a simple argument \cite{gid,knthree}
that the pion case should be similar, and involves the radion (position 
of the IR brane) as pion dual. Heisenberg uses for the nonlinear scalar pion 
action the DBI action
\be
S=l^{-4}\int \sqrt{1+l^4[(\partial \phi)^2+m^2\phi^2]}
\ee
which is just the radion action at $m=0$ and may provide a nonlinear 
version of the stabilized radion at $m\neq 0$. He then collides 
pion field shockwaves, which at $x^+\leq 0, x^-\leq 0$ are just
\be
\phi = \phi_1 (x^i) \delta(x^+)+\phi_2 (x^i)\delta (x^-)
\ee  

The field $\Phi$ in (\ref{shock}) takes the role of the pion field $\phi$ in 
the Heisenberg picture, with exponential decay $\Phi\sim r^{-1/2} e^{-M_1r}$
similar to the pion field decay around hadrons. 
The exact mapping of the dual scattering
to Heisenberg scattering suggests that there should be a nonlinear pion 
field soliton being created in the collision, mapped to the black hole 
created in the dual collision. Again, strictly speaking this applies only 
for the lightest glueball case, but it should also be extended to the pion 
case by the argument in \cite{gid,knthree}.

One cannot find this soliton in perturbation theory, and Heisenberg 
calculated only the perturbative solution close to the interaction, 
and far away, where one has just free waves (free pions, radiated away). 
In the gravity dual we could argue for black hole formation because 
we could calculate the formation of a trapped surface at the interaction
point, using a formalism developped for flat d=4 in \cite{eg} and extended 
to curved general d in \cite{kn}. The beauty of GR is that we have a theorem 
stating that a horizon will form outside the trapped surface.

In \cite{nastase2}, we have analyzed in detail the energy regimes of 
gauge theories and found that before the Froissart saturation, one 
has a further energy regime. The gravity dual scattering happens at a certain 
distance from the IR brane, and the black hole
starts to feel the AdS size before reaching the IR brane. Scattering in 
$AdS_{d+1}\times X_{\bar{d}}$ gives a field $\Phi $ (solution to the 
Poisson equation (\ref{poisson}))
\be
\Phi\sim \frac{1}{r^{2(d-1)+\bar{d}}}=\frac{1}{r^{11}}
\ee
and gives a gauge theory cross section $\sigma_{gauge}\sim s^{1/11}$. 
We have argued that this behaviour should onset at the scale $\hat{E}_R
= N^2 \Lambda_{QCD}$, and the maximal Froissart behaviour should onset 
at an unknown energy scale $\hat{E}_F$, depending on the IR details of the 
gravity dual, reached when the black hole horizon merges with the IR brane. 
In the case when the pion is the lightest excitation, the Froissart 
behaviour would onset at a scale $\hat{E}_F'(m_{\pi})<\hat{E}_F$, reached 
when the IR brane bending engulfs the black hole. 

We have also argued that this analysis goes through for real QCD as well, 
as string $g_s$ and $\alpha '$ corrections to the scattering 
\cite{kn} are small above ${E}_R$, 
and then so are $1/N$ and $1/(g^2_{YM}N)$ corrections in the gauge theory
above $\hat{E}_R$. In QCD $\hat{E}_R= N_c^2 M_{1,glueball}\sim 10 GeV$, 
and thus above it one should see the ``soft Pomeron'' behaviour $\sigma
_{tot}(s)\sim s^{1/11}$, consistent with experimental evidence. 

We noticed that the Heisenberg model can be mapped to the dual A-S 
shockwave scattering picture even before the saturation of the Froissart 
bound. All we have to do is relax Heisenberg's assumption of exponentially 
decaying pion wavefunction around the hadrons, and have instead the 
wavefunction $\Phi$ of the A-S shockwave which is mapped to a 
wavefunction of the lightest glueball field. In this case however, 
unlike for the maximal Froissart behaviour, the dual 
picture is not 4 dimensional anymore, it is 10 dimensional, and we have 
the usual holography. This Heisenberg description of the 
gauge theory scattering means that 
the black hole being formed in the dual scattering is still mapped to a 
nonlinear effective field soliton, but the effective field is the 
lightest glueball. Above $\hat{E}_R\sim 10 GeV$, this description 
is exact, below it could be modified due to string corrections. 

So we predict a nonlinear effective field soliton, dual to a black hole,
being produced in 
high energy collisions, at least above 10GeV, and maybe also above 
$\hat{M}_P= N_c^{1/4}M_{1, glueball}\sim 1-2GeV$, but in the latter case 
the soliton is small enough so it could be hard to isolate. 
So where is this soliton at higher energies? Then clearly it should have 
been already observed, as its size should be sufficiently large.

We argue that it has already been observed, at RHIC. Indeed, RHIC functions 
at a center of mass energy of 100+100 $GeV$ per nucleon, in the collision 
of Au+Au. This energy should be enough even for the formation of a soliton
in the collision of just two nucleons, but certainly for two Au nuclei (A=197
for the common isotope). At that huge total energy, one will clearly be in 
the maximal Froissart regime in the gauge theory, so one will create the 
pion analog of black holes on the IR brane. But RHIC does observe a fireball
at finite temperature, exactly as expected! 

RHIC expected to see a ball of ``Quark Gluon Plasma'' 
(QGP), that is a free gas 
of deconfined quarks and gluons, at a temperature above the phase transition 
temperature for
 chiral restoration (and maybe deconfinement), which is obtained from the 
lattice to be in the $170-180 MeV$ domain. One does see a thermal object 
indeed, but at its core is a deconfined, yet strongly coupled phase dubbed
``Color Glass Condensate'' (CGC). As the CGC expands, it is found to go 
into the thermal QGP, and eventually decay into free pions (and some other 
particles), thermally distributed corresponding to a ``freeze-out'' temperature
of $176 MeV$ \cite{ll,bm}. There seems to be some debate about whether the 
CGC refers to the colliding objects or the object formed in the collision, 
but we will adopt the terminology of the latter case in the following.

We want to argue that this is just the nonlinear pion field soliton, dual 
to the black hole on the IR brane, and the ``freeze-out''
temperature observed is the black hole temperature. As we mentioned, the 
nonlinear field soliton will decay into free pions, dual to the 
black hole radiating away gravitons. Let us therefore 
try to calculate the temperature.

We will try to follow the usual 4d black hole thermodynamics calculations. 
We start from the thermodynamics relation \cite{bch}
$d{\cal M}= TdS$, where ${\cal M}$
is the black hole mass, T is temperature and S its entropy. We assume that
we still have $S= M_{P,4}^2 A/4$, where A is the horizon area. 
This relation was 
derived by knowing the temperature of the black hole, but it seems to be a 
fundamental relation in quantum gravity, and has been tested in numerous
string theory calculations \cite{sv,cm},...(see e.g. \cite{maldacena}
for more references). Then, as the black hole on the IR brane is 
spherically symmetric, we have $A=4\pi r_H^2$, therefore ($l_H=2r_H$)
\be
d{\cal M}= \frac{\pi T}{4} M_{P,4}^2d(l_H^2)
\ee
We notice that if we have 
\be
M_{P,4}^2 d(l_H^2)=  \frac{d{\cal M}}{a M_1}
\label{horiz}
\ee
with $M_1$ the mass of the KK graviton (mapped to the pion mass) and $a$ 
a numerical constant, we get 
\be
T=a\frac{4M_1}{\pi}
\label{temp}
\ee
Let's see what can we say in perturbation theory, without knowing the full 
nonperturbative black hole solution. In perturbation theory \cite{gid}
\be
-g_{00}=1- \frac{k G_4 {\cal M}}{r} e^{-M_1r}
\label{pert}
\ee
where k is a numerical constant and $G_4\equiv (RM_P^3)^{-1}$. 
Then 
\be
r_H\simeq \frac{1}{M_1}ln (k G_4 {\cal M} M_1)
\ee
that implies
\be
M_{P,4}^2 d(r_H^2)= \frac{d{\cal M}}{M_1}2(\frac{M_{P,4}^2}{M_1{\cal M}})
ln\; k(\frac{ M_1{\cal M}}{M_{P,4}^2})  
\ee
which is different than (\ref{horiz}), but not too much if ${\cal M} M_1 
\sim M_{P,4}^2$. In any case, this is perturbation theory, and there is 
no reason to trust it near the horizon. 

Let's try a different perspective and use Hawking's original calculation of 
the temperature of 4d black holes of Kerr-Newman type (with mass, charge
and angular momentum) \cite{hawking}. He found the expression
\be
T=\frac{\kappa}{2 \pi}
\label{hawking}
\ee
where $\kappa$ is the surface gravity of the horizon, defined by 
\be
\nabla^a (\chi^b \chi_b)=-2\kappa \chi^a
\ee
and $\chi^a$ is the Killing vector that is tangent to the null geodesic 
generator of the horizon, in our case just $\chi^a =(\partial/\partial t)^a$.
Then
\be
4\kappa ^2 = \lim_{horizon} \frac{||\nabla^a (\chi_b \chi^b)||^2}{
\chi^a \chi_a}= 
\lim_{horizon} \frac{[\partial_r (g_{tt})]^2 g^{rr}+ [\partial_y (g_{tt})]^2 
g^{yy}}{g_{tt}}
\label{calc}
\ee
For a Schwarzschild black hole, $g^{rr}= -g_{tt}= 1- 2{\cal M}G_4/r$, 
and $\partial_r (g_{tt})|_{horizon}= 1/(2{\cal M}G_4)$, giving $T= 1/(8\pi
{\cal M}G_4)$.

But in general we have $g^{rr}=g_{tt}$ for a black hole, in fact that is 
how one nonextremalizes D-brane solutions in flat space 
\cite{ct} to obtain black branes solutions, so we will 
assume that the same holds now, at least at the horizon. 
Also, for our solution, the horizon 
is almost 4 dimensional, so $g^{yy}\sim 1$ and the second term in (\ref{calc})
can be neglected at the horizon, thus we get 
\be
2\kappa = | \partial_r(g_{tt})|_{horizon}
\ee
Assuming that near the horizon $g_{tt}$ looks similar to the perturbative
solution (\ref{pert}), namely dominated by an exponential decay,
with subleading power law behaviour,  of the type
\be
g_{tt}\simeq  1-\frac{Ae^{-M_1 r}}{r^n}
\ee
we get 
\be
\kappa =\frac{1}{2}
|\partial_r (g_{tt})|_{horizon}\simeq \frac{1}{2}(M_1+\frac{n}{r_H})\simeq M_1
 \ee
If we also have the law (\ref{hawking}), we get 
\be
T\simeq \frac{M_1}{2\pi}
\ee
which is indeed of the form in (\ref{temp}), with $a\simeq 1/8$. 
The calculation
of the surface gravity of the horizon was fairly general, the only 
possible weak point being the assumption $g^{rr}=g_{tt}$ at the horizon.
But there is no a priori reason to still have exactly (\ref{hawking}), 
as that was done for black holes in asymptotically flat 4d space, and now 
we have a black hole on the IR brane, for highly curved AdS space,
so the topological structure of infinity, thus the Penrose diagram 
used for the calculation, could be modified. One 
should redo Hawking's analysis in that background. 

Thus we have a reason to believe that (\ref{temp}) is correct. It also 
makes sense from the point of view of dimensional analysis. The temperature
is independent of ${\cal M}$ and $M_{P,4}$, and we understand this physically
since the black hole is dominated by the exponential decay, whose only scale 
is $M_1$. The solution $a\simeq 1/8$ is probably just due to our use of the 
formula (\ref{hawking}), so we leave it a free parameter.

Indeed, if $a=1$, and we replace $M_1$ by the pion mass $m_{\pi}$, by which 
we mean of course the average mass $(m_{\pi^+}+m_{\pi^-}+m_{\pi^0})/3$, we get 
\be
T=\frac{4}{\pi}<m_{\pi}>= 175.76 MeV
\label{tempe}
\ee
remarkably close to the experimental value of the RHIC fireball ``freeze-out''
of $176 MeV$ \cite{ll,bm}. Since that value is argued to be also the 
temperature of the chiral restoration (and maybe deconfinement) 
phase transition (lattice results also obtain a value in that 
region), it is possible that the dual calculation gives also the temperature
of that phase transition. 

For more experimental evidence of the identification of the fireball and 
the dual black hole, we turn to the calculation of the quantity 
shear viscosity over entropy, $\eta/s$. It was argued \cite{kss} that this 
quantity is always bounded from below by $1/(4\pi)$ 
(see also the $\eta$ calculation in \cite{pss} and \cite{kss2}). 
In \cite{bl} (see also \cite{kss2,buchel}) a theorem was 
proven stating that black holes in type IIB gravity duals saturate 
this bound, thus $\eta / s= 1/(4\pi)$. But experimentally, 
under the assumption of a hydrodynamic model (about which there is 
still considerable debate), one finds 
that the fireball has a value that is very close to it, 
$\eta / s\sim 0.1-0.3$ \cite{shuryak,shuryak2}.
What was regarded as an interesting curiosity can now be argued should 
be rigorous equality, as the fireball should be nothing but a 
gravity dual black hole. 

Therefore we can confidently say that the RHIC fireball is the conjectured
nonlinear pion field soliton dual to the black hole on the IR brane. 
Let us explore the consequences of this. The first and most obvious is the 
evolution of the fireball. One first creates a pion field soliton=black hole,
that 
quickly thermalizes and then rapidly decays through the emission of 
pions= dual gravitons, almost in an explosion.  Thus the ``Color Glass 
Condensate'' (CGC) state at the core of the fireball is the pion field 
soliton=black hole interior. 

One of the experimental reasons for saying that the state at the fireball 
core is a glass-like state (CGC), is the fact that there are almost no hard 
scattering events, a phenomenon known as ``jet quenching'' (absence of
jets). Hard scattering signals the presence of perturbative QCD interactions
(due to asymptotic freedom). The fact that these hard scattering events are 
suppressed is strange, and would usually mean that the large 
transverse momentum of the perturbative interaction is somehow 
atenuated (absorbed) over the very short distance of the fireball. 
This seems impossible in perturbative QCD, hence the QGP picture was 
replaced by an unknown CGC picture, implying very strong interactions
that absorb the transverse momentum. 

Now we have a very simple picture of the CGC interaction. 
Any particles that are in the 
interaction region get caught in the formation of the black hole, and 
can only escape by being radiated away in a thermal manner. Thus 
``jet quenching'' is nothing other than the usual black hole ``information
paradox'', that information gets inside the black hole and only radiation 
gets out. The idea that jet quenching can be explained by matter falling 
into a black hole by AdS-CFT has been also proposed in the context of 
${\cal N}=4$ SYM at finite temperature in \cite{sz}, but we are treating here
the case of scattering in the real QCD gravity dual (and our 4d space is at 
zero temperature).

That also means that it will probably be very hard, if not impossible 
to probe the interior of the fireball with extra energetic particles 
present close to the interaction region that enter the fireball.
 That would amount in the black hole picture to 
probing the interior of the black hole with other particles, and is 
clearly impossible. However, given that the formed black hole is small 
enough and its lifetime finite, it could be possible in principle, by 
the right combination of time and length scales, to probe the formation 
and decay phases of the black hole. 

Finally, what kind of state is the CGC, meaning how do particles inside it 
interact? The question is clearly answered in both the effective field theory
and in the dual picture. In the effective field theory, we have a pion field 
soliton, thus particles should interact mostly by exchange of pions. As we are
at distances $r<1/m_{\pi}$, the pion Yukawa potential is replaced by the 
Coulomb potential. In the dual picture, we have the perturbative gravitational
(Newton) potential (\ref{pert}), thus at distances $r<1/M_1$ we have 
\be
U= -\frac{k G_4 {\cal M}}{r} 
\ee

As we mentioned, the actual dual to pion interaction should be brane 
bending (radion interaction), for which one has \cite{gid,nastase2}
\be
U\sim \frac{\delta L}{L}|_{lin} \sim -\frac{k G_4 (M_LR){\cal M}}{r}
\ee
and correspondingly the gauge theory potential energy will be
($\hat{M}_{P,4}= N^{3/8}M_1, M_L\rightarrow m_{\pi}$)
\be
V(r) =-\frac{m_{\pi} M_1^{-1}}{N_c^{3/4}M_1^2}\frac{k{\cal M}_1
{\cal M}_2}{r}= -0.06 GeV^{-2}
(M_1[GeV])^{-3}
\frac{k{\cal M}_1{\cal M}_2}{r}
\label{potential}
\ee
where $M_1=M_{1, glueball}$ and we have left the 
numerical constant k free since it depends on the exact model of how 
brane bending and gravity interact in the dual, and 
anyway we expect that the relation of 
energy scales like $M_P$ to gauge theory quantities could be renormalized 
by string corrections. Here ${\cal M}$ is the (relativistic)
mass of the corresponding particle. 

However, we should note now that we have done something not obviously 
right. We have extended the regime of validity of the Newtonian approximation
to the interior of the black hole, where we have no justification for this 
approximation. But remember that this is not a usual black hole, as it 
corresponds to a KK type of gravity, with mass $M_1$. Also, we are talking 
now about the interaction of two massive perturbations inside the black 
hole, and not about the background gravitational field. If the black 
hole is large enough, the interaction of the two masses should still be 
Newtonian.
This is clearer in the 
dual picture where we have pion fields, and pion exchange should dominate
the interactions
between particles. 

Also note that usual black holes have singularities, but 
the black holes created in this collision most likely will not. Indeed, 
for one thing it would be hard to imagine the brane bending analog of the 
black hole singularity: it would be a pinch-like singularity that seems 
unphysical. Also, in the field theory, a pion field singularity seems 
very unlikely. It could also be the case that these 
black holes are created and decay quickly, and the singularity cannot form. 

In any case, it seems that the interior of the formed IR brane black hole 
should be smoothed out with respect to the case of usual (Schwarzschild)
black holes.

Thus quarks and gluons will be deconfined in the CGC (pion field soliton), 
and will interact with the Coulomb-like (or rather, 
Newton-like) potential (\ref{potential}), which can create bound 
states, similar to atomic states (or rather, ``solar systems''), that 
was argued that could a priori be experimentally detected \cite{shuz}. 

Note that in this paper we have analyzed collisions at RHIC, which should be 
in the maximal Froissart regime, but even for collisions at 
$\hat{E}_R\sim 10 GeV < \sqrt{s}< \hat{E}_F'$ one will create black holes 
in the gravity dual, and solitons of the effective lightest glueball field 
in the gauge theory. This (much smaller) soliton will also decay, producing 
an (almost thermal) distribution, so one should find evidence for the 
thermalization of the decay products. Its temperature however will be 
different, as that temperature depends on the gravity dual, 
possibly on $\sqrt{s}$, and on also the 
mass $M_{1, glueball}$. 

In conclusion, we have seen that the observed RHIC fireball is just 
(the pion analog of) a gravity dual black hole. We have seen that 
the unknown CGC state at the middle of the fireball is the interior of 
the black hole, and particles inside it interact with a Newtonian 
potential. The horizon of the black hole is the limiting (``freeze-out'')
surface of the pion field soliton, which emits radiation at a temperature
given by (if the nonperturbative constant a=1) (\ref{tempe}), very close 
to the experimental value of 176 MeV \cite{ll,bm}. Most likely there will be 
no singularity for this black hole. 

Then RHIC is really a string theory testing machine, that analyzes the 
formation and decay of dual black holes, and giving information about the 
black hole interior. Probing the formation and decay with external particles
seems hard, but a priori not impossible. The same applies for the formation
of bound states of the potential (\ref{potential}).

{\bf Acknowledgements} I would like to thank Kyungsik Kang and Antal Jevicki
for discussions. I would also like to thank the organizers of the ITP
Santa Barbara ``QCD and string theory'' conference in November 2004, during 
which I learned about the exciting physics being done at RHIC. I would 
like to thank E. Shuryak and R. Pisarski for pointing out to me  
references about RHIC physics.
This research was  supported in part by DOE
grant DE-FE0291ER40688-Task A.

\end{document}